\documentclass[twocolumn,secnumarabic,amssymb, amsmath, nofootinbib,tightenlines, nobibnotes, color, aps, prl,epsfig]{revtex4}
\usepackage{graphicx}
\usepackage{dcolumn}
\usepackage{bm}
\begin{document}
\preprint{APS/123-QED}
\title{ Geometrical scaling of heavy-quark contributions in the low $x$ region }

\author{G.R.Boroun}%
 \email{boroun@razi.ac.ir }
\affiliation{ Department of Physics, Razi University, Kermanshah
67149, Iran}

\date{\today}
\begin{abstract}
We describe the determination of the heavy quarks structure
functions $F_{2,L}^{\mathcal{Q}\overline{\mathcal{Q}}}$  with help
of the scaling properties. We observe that the structure functions
for inclusive charm and bottom production exhibits geometric
scaling  at low $x$. The geometrical scaling means that the heavy
quark  structure function is a function of only one dimensionless
variable $\tau{\equiv}Q^{2}/Q^{2}_{sat}(x)$ including quark mass.
These results are valid for any value of $\delta$, being
$x^{-\delta}$ the behavior of the parton densities at low $x$. The
determination of the heavy quark structure function is presented
as a parameterization of the proton structure function
$F_{2}(x,Q^{2})$ and its derivative. Analytical expressions for
$F_{2,L}^{\mathcal{Q}\overline{\mathcal{Q}}}$ in terms of the
effective parameters of the parameterization of $F_{2}(x,Q^{2})$
,with respect to the BDH and ASW models, are presented. To study
the heavy quark production processes, we use the collinear results
in DAS approach. Numerical calculations and comparison with HERA
data demonstrate that the suggested method provides reliable
$F_{2}^{\mathcal{Q}\overline{\mathcal{Q}}}$, $R^{c\overline{c}}$
and $\sigma^{\mathcal{Q}\overline{\mathcal{Q}}}$ at low $x$ in a
wide range of the low absolute four-momentum transfers squared
($4~\mathrm{GeV}^{2}<Q^{2}<2000~\mathrm{GeV}^{2}$). Also, in the
HERA kinematic range, the ratio of
$F_{2}^{c\overline{c}}/F_{2}^{b\overline{b}}$,
$F_{2}^{c\overline{c}}/F_{2}^{BDHM}$ and
$F_{2}^{b\overline{b}}/F_{2}^{BDHM}$ are obtained. Expanding the
method to low and ultra low values of $x$ can be considered in the
process analysis of the LHeC and FCC-eh colliders.
\end{abstract}
 \pacs{***}
\maketitle
\subsection{1. INTRODUCTION}

In earlier HERA experiments, data on charm and bottom structure
functions, $F_{2}^{c\overline{c}}(x,Q^{2})$ and
$F_{2}^{b\overline{b}}(x,Q^{2})$, were found to be around $\%25$
and $\%1$ of the proton structure function $F_{2}(x,Q^{2})$ at
small $x$ respectively. In ultrahigh energy processes, at
extremely small $x$, these percents will be increased and checked
in high energy processes such as the Large Hadron electron
Collider (LHeC) and the Future Circular Collider electron-hadron
(FCC-eh) projects which run to beyond a TeV in center-of-mass
energy [1]. These new colliders  lead into the region of high
parton densities at low $x$ as the kinematic reach of maximum
$Q^{2}{\simeq}1~\mathrm{TeV}^{2}$ and $x{\simeq}10^{-5..-6}$ for
LHeC and $10^{-7}$ for FCC-eh. By electron-proton (ep) colliders,
the heavy quarks can be produced in pair in the deep inelastic
scattering (DIS) through neutral current (NC) production. In these
regions, the gluon density becomes predominant and the heavy
quarks behavior will be checked in the photo-gluon fusion
reactions. The production of heavy quarks at HERA depend on the
mass of these quarks and thus the calculations of cross sections
depend on a wide range of perturbative scales $\mu^{2}$. The first
production of heavy quarks at HERA is keeping by fixed number of
parton densities (fixed flavour number schemes, FFNS) close to
threshold $\mu^{2}{\sim}m^{2}_{\mathcal{Q}}$ (where
$m_{\mathcal{Q}}$ is the heavy quark mass), as
\begin{eqnarray}
F^{\mathcal{Q}\overline{\mathcal{Q}}}
_{2,L}|_{\mathrm{FFNS}}=\sum_{j=0}^{\infty}a_{s}^{j}(n_{f})\sum_{i=q,g}
H_{k,i}^{(j)}(n_{f}){\otimes}f_{i}(n_{f}),
\end{eqnarray}
where $H^{,}s$ are the Wilson coefficients for the DIS heavy-quark
production [2] and the dynamics of flavor-singlet quark and gluon
distribution functions, $q^{s}$ and $g$, are defined by
\begin{eqnarray}
q^{s}(x,n_{f},\mu^{2})&=&\sum_{l=1}^{n_{f}}[f_{l}(x,n_{f},\mu^{2})+\overline{f}_{l}(x,n_{f},\mu^{2})],\nonumber\\
g(x,n_{f},\mu^{2})&=&f_{g}(x,n_{f},\mu^{2}),
\end{eqnarray}
where $n_{f}$ is the number of active quark flavors. FFNS can be
used on the threshold of $\mu^{2}{\sim}m^{2}_{\mathcal{Q}}$ and
for $\mu^{2}{>}m^{2}_{\mathcal{Q}}$ the variable-flavour-number
scheme (VFNS) is used where the treatment of resummation of
collinear logarithms $\ln(\mu^{2}/m^{2}_{\mathcal{Q}})$ is
achieved. For realistic kinematics it has to be extended to the
case of a general- mass VFNS (GM-VFNS) which is defined similarly
to the zero-mass VFNS (ZM-VFNS) in the $Q^{2}/m^{2}_{\mathcal{Q}}$
limit [3-6]. The heavy-quark structure functions derived  using
the zero-mass VFN scheme (ZM-VFNS) as
\begin{eqnarray}
F^{\mathcal{Q}\overline{\mathcal{Q}}}
_{2,L}|_{\mathrm{ZM-VFNS}}&=&\sum_{j=0}^{\infty}a_{s}^{j}(n_{f}+1)\sum_{i=q,g,\mathcal{Q}}
C_{k,i}^{(j)}(n_{f}+1)\nonumber\\
&&{\otimes}f_{i}(n_{f}+1),
\end{eqnarray}
where $C^{,}s $ are the Wilson coefficients at the $j$-th order
and  $a_{s}=\frac{\alpha_{s}}{4\pi}$ is the QCD running coupling.
For $Q^{2}{\simeq}m_{\mathcal{Q}}^{2}$ VFNS is valid where it
includes a combination of the ZM-VFNS with FFNS.\\
The heavy-quark structure functions obtained in DIS at HERA, from
the measurement of the inclusive heavy quark cross sections, are
an important test of the QCD [7]. The heavy-quark reduced cross
section defined in terms of the heavy-quark structure functions as
\begin{eqnarray}
\sigma^{\mathcal{Q}\overline{\mathcal{Q}}}_{\mathrm{red}}(x,Q^{2})&=&
F_{2}^{\mathcal{Q}\overline{\mathcal{Q}}}(x,Q^{2})-f(y)F_{L}^{\mathcal{Q}\overline{\mathcal{Q}}}(x,Q^{2}),
\end{eqnarray}
where $f(y)=\frac{y^{2}}{1+(1-y)^{2}}$ and $y=Q^{2}/sx$ is the
inelasticity with s the ep center of mass energy squared. In HERA
kinematic range the contribution
$F_{L}^{\mathcal{Q}\overline{\mathcal{Q}}}$ is small, therefore
the heavy-quark structure function
$F_{2}^{\mathcal{Q}\overline{\mathcal{Q}}}$ is obtained from the
measured heavy-quark cross section. Future circular colliders
(LHeC and FCC-he) will extend the ratio
$F_{L}^{\mathcal{Q}\overline{\mathcal{Q}}}/F_{2}^{\mathcal{Q}\overline{\mathcal{Q}}}$
into a region of much smaller $x$ and higher $Q^2$. Indeed these
new colliders are the ideal place to resolve this ratio [1].\\
In deep inelastic scatting (DIS), the photon-proton cross section
for heavy-quark production is related to the heavy-quark structure
function with the following form
\begin{eqnarray}
\sigma^{\mathcal{Q}\overline{\mathcal{Q}}}(x,Q^{2})=4\pi^{2}
\alpha_{em}
F_{2}^{\mathcal{Q}\overline{\mathcal{Q}}}(x,Q^{2})/Q^{2}.
\end{eqnarray}
One of the well-known property of deep inelastic scatting  at low
$x$ is geometrical scaling (GS) [8-10]. GS is the statement that
the $\sigma^{\mathcal{Q}\overline{\mathcal{Q}}}(x,Q^{2})$, which
is a priori function of two independent variables $(x,Q^{2})$,
depends only on a specific combination of them,
$\tau=Q^{2}/Q^{2}_{s}(x)$, where function $Q_{s}(x)$ called
saturation scale. In principle
$\sigma^{\mathcal{Q}\overline{\mathcal{Q}}}(x,Q^{2})=\sigma^{\mathcal{Q}\overline{\mathcal{Q}}}(\tau)$.
The saturation scale is a border between dense and dilute gluonic
systems where was taken in the following form
\begin{eqnarray}
Q^{2}_{s}(x)=Q^{2}_{0}(x/x_{0})^{-\lambda}
\end{eqnarray}
Here $Q_{0}=1~\mathrm{GeV}$ and the parameters $x_{0}$ and
$\lambda$ were determined from a fit to DIS data at low $x$ [11].
The gluon saturation appears formally due to the nonlinearities of
parton evolution at low $x$ given by JIMWLK equations [12]. A
description of this low $x$ regime is given by the Color Glass
Condensate (CGC), a semi-classical effective field theory (EFT)
for low $x$ gluons. In this model the transition between the color
transparency and the dipole correlator is delineated by defining
the saturation scale $Q_{s}(x)$. These two regimes are defined
into small separations $r_{\bot}$ and large distances
$r_{\bot}$ respectively, where $r_{\bot}$ is the dipole separation [13-14].\\
In heavy quarks production GS is expected to be violated due to
large quarks mass. To do this we replace the Bjorken scaling by
the rescaled variable
\begin{eqnarray}
\chi=x(1+\frac{4m^{2}_{\mathcal{Q}}}{Q^{2}})=\frac{Q^{2}+4m^{2}_{\mathcal{Q}}}{W^{2}}
\end{eqnarray}
and keep $m_{\mathcal{Q}}{\neq}0$. Here $W$ is the total energy of
the $\gamma^{*}p$ system. The rescaling variable is one of the
ingredients used in the GM-VFNS. The cross section for heavy quark
pair production vanishes when $\chi$ goes to 1 for $W$ as small as
$2m_{\mathcal{Q}}$ [15]. Main purpose of this paper is to asses
the GS in heavy quark pair production. To this end we take into
account $m_{\mathcal{Q}}$ in
\begin{eqnarray}
\tau=Q^{2}/Q^{2}_{0}(1+\frac{4m^{2}_{\mathcal{Q}}}{Q^{2}})^{\lambda}
(x/x_{0})^{\lambda},
\end{eqnarray}
to H1 and ZEUS data [16] on charm and bottom
structure functions.\\
In the present paper we present a method of extraction of the
heavy quark structure functions,
$F^{\mathcal{Q}\overline{\mathcal{Q}}}_{2}(x,Q^{2})$ and
$F^{\mathcal{Q}\overline{\mathcal{Q}}}_{L}(x,Q^{2})$ in the
kinematical region of low values of the Bjorken variable $x$ from
the parameterization models by relying on the saturation scaling
arguments. Indeed a simple parametrization for the heavy quark
structure function in the region of $x {<}0.1$ in a wide interval
of photon virtualities is proposed. The organization of this paper
is as follows. In section II we introduce the basic formula used
for the definition of evolution equations and color dipole
picture. In section III we present the heavy quark structure
functions with respect to the proton parameterization models. The
main results and finding of the present heavy quark structure
functions due to the modified geometrical scaling is discussed in
detail in section IV. In the same section, the feasibility of
measuring $F^{\mathcal{Q}\overline{\mathcal{Q}}}_{2} (x,Q^{2})$,
in the available energy to the range of new collider energies
(i.e., LHeC and FCC-eh), is investigated. Conclusions and
summary are summarized on Sec.V.\\

\subsection{2. Deep-inelastic Structure Functions}

Measurements at HERA have shown that heavy flavor production in
DIS proceeds predominantly via the photon-gluon fusion process
$\gamma{g}{\rightarrow}\mathcal{Q}\overline{\mathcal{Q}}$, as the
reaction under study is
\begin{eqnarray}
e^{-}+P{\rightarrow}~e^{-}+\mathcal{Q}\overline{\mathcal{Q}}+X,
\end{eqnarray}
where $P$ is a proton, $\mathcal{Q}\overline{\mathcal{Q}}$ is a
pair heavy-quark and $X$ is any hadronic state allowed. The
differential cross section
$\sigma^{\mathcal{Q}\overline{\mathcal{Q}}}$(hereafter
$\mathcal{Q}=c,b$) can be presented in the following form
\begin{eqnarray}
\frac{d^{2}\sigma^{\mathcal{Q}\overline{\mathcal{Q}}}}{dxdQ^{2}}&=&
\frac{2\pi
\alpha^{2}}{xQ^{4}}[(1+(1-y)^{2})F_{2}^{\mathcal{Q}\overline{\mathcal{Q}}}(x,Q^{2})\nonumber\\
&&-\frac{y^{2}}{2}F_{L}^{\mathcal{Q}\overline{\mathcal{Q}}}(x,Q^{2})],
\end{eqnarray}
where $F_{k}^{\mathcal{Q}\overline{\mathcal{Q}}}(x,Q^{2})$
(hereafter $k = 2, L$) is heavy quark parts of the proton
structure function $F_{k}(x,Q^{2})$ [17,18]. Analogous studies
have been performed in Ref.[18] which is based on the transverse
momentum dependent (TMD) gluon density. TMD, or non-integrated,
functions depending on the fraction of the longitudinal momentum
$x$ of the proton carried by the parton, the two-dimensional
transverse momentum of the parton $\mathbf{K}_{T}^{2}$ , and the
hard scale $\mu^{2}$ of a complex process, contain nonperturbative
information about the proton structure, including the transverse
momentum.\\
In the small $x$ region, the heavy quark structure functions  in
the collinear generalized double asymptotic scaling (DAS) approach
are given by [18]
\begin{eqnarray}
F_{k}^{\mathcal{Q}\overline{\mathcal{Q}}}(x,Q^{2})=C_{k,g}(x,Q^{2},m^{2}_{\mathcal{Q}}){\otimes}
f_{g}(x,\mu^{2}),
\end{eqnarray}
where $C_{k,g}(x,Q^{2})$ are the Wilson coefficient functions and
$\mu$ is the renormalization scale. The symbol $\otimes$ is the
Mellin convolution
\begin{eqnarray}
F_{k}^{\mathcal{Q}\overline{\mathcal{Q}}}(x,Q^{2})=\int_{x}^{x_{2}}\frac{dy}{y}
C_{k,g}(y,\xi)f_{g}(\frac{x}{y},\mu^{2}),
\end{eqnarray}
where $x_{2}=1/(1+4\xi)$, $\xi=m^{2}_{\mathcal{Q}}/Q^{2}$ and
$f_{g}(x,Q^{2})$ is the gluon density. The renormalization and
factorization scales were set to be equal to
$\mu^{2}_{R}=4m^{2}_{\mathcal{Q}}+Q^{2}$ and $\mu^{2}_{F}=Q^{2}$,
respectively [19,20]. The coefficient functions at leading order
(LO) up to  next-to-next-to leading order (NNLO) approximations
are defined [18,21]
\begin{eqnarray}
C_{k,g}(x,a_{s})&=&e^{2}_{\mathcal{Q}}\sum_{n=0}(a_{s}(\mu^{2}))^{n+1}B^{(n)}_{k,g}(x,\xi),
\end{eqnarray}
where in the high energy regime the coefficients
$B^{(n)}_{k,g}(x,\xi)$ have the compact forms defined in
literature and $n$ denotes the order in running coupling
$\alpha_{s}(\mu^{2})$ [18-21].\\
Using the fact that the non-singlet contribution can be ignored
safely at low values of $x$. According to the famous DGLAP
equations, the evolution of the singlet structure function reads
\begin{eqnarray}
\frac{{\partial}F_{2}(x,Q^{2})}{{\partial}{\ln}Q^{2}}&=&-\frac{1}{2}\sum_{n=0}(a_{s}(Q^{2}))^{n+1}[
\widetilde{P}^{(n)}_{ss}(x)
{\otimes}F_{2}(x,Q^{2})\nonumber\\
&&+<e^{2}>\widetilde{P}^{(n)}_{sg}(x){\otimes}xf_{g}(x,Q^{2})],
\end{eqnarray}
where
\begin{eqnarray}
\widetilde{P}_{ab}^{(0)}(x)&=&{P}_{ab}^{(0)}(x),\nonumber\\
\widetilde{P}_{ab}^{(n)}(x)&=&{P}_{ab}^{(n)}(x)+\sum_{m}B^{(n)}_{2,b}(x){\otimes}
{P}_{ab}^{(n-1)}(x)\nonumber\\
&&+2\beta_{0}\sum_{m}B^{(n)}_{2,b}(x){\otimes}\delta(1-x)
+...,~n>0 \nonumber
\end{eqnarray}
The quantities $\widetilde{P}_{ab}$$^{,}s$ are expressed via the
known splitting  and Wilson coefficient functions  in literatures
 and $P_{ab}$ are the splitting functions in [18-21]. In the above equation $<e^{k}>$ is
the average of the charge $e^{k}$ for the active quark flavors,
$<e^{k}>=n_{f}^{-1}\sum_{i=1}^{n_{f}}e_{i}^{k}$.\\
Parametrization of the proton structure function suggested in
Ref.[22] by M.M.Block, L.Durand and P.Ha, in what follows referred
to as the MDH model (BDHM), describe fairly well the available
experimental data on the reduced cross sections and, at
asymptotically low $x$, provide a behavior of the cross sections
$\sim\ln^{2}1/x$, in an agreement with the Froissart predictions.
This suggested method provides reliable proton structure functions
$F_{2}(x,Q^{2})$ with $x{\leq}0.1$ in a wide range of the momentum
transfer $(0.15~\mathrm{GeV}^{2} < Q^{2} < 3000~\mathrm{GeV}^{2})$
and can be applied as well in analyses of ultrahigh energy
processes with cosmic neutrinos. The explicit expression for the
BDHM reads
\begin{eqnarray}
F_{2}^{\mathrm{BDH}}(x,Q^{2})=D(Q^{2})(1-x)^{\nu}\sum_{m=0}^{2}A_{m}(Q^{2})L^{m},
\end{eqnarray}
where the effective parameters are defined in Refs.[22,23]. In the
framework of color glass condensate an analytic proton structure
function including quark mass has suggested by authors in
Ref.[24]. Those kept the quark mass in photon wave function and
obtained the massive proton structure function. The proton
structure function in the dipole frame, obtained via
\begin{eqnarray}
F_{2}(x,Q^{2})&=&\frac{Q^{2}}{4\pi^{2}\alpha}[\sigma_{L}^{\gamma^{*}p}(x,Q^{2})+\sigma_{T}^{\gamma^{*}p}(x,Q^{2})]
\end{eqnarray}
where the subscripts $L$ and $T$ denote the longitudinal and
transverse polarizations of the virtual photon. In this frame the
photon undergoes the hadronic interaction through dissociation of
the photon into a quark-antiquark $(q\overline{q})$ pair of flavor
$f$ and size $r$, called dipole, and the interaction of the
quark-antiquark pair with the target proton [25]. Within the
dipole framework of the $\gamma^{*}p$ scattering the
photoabsorpion cross section is written as
\begin{eqnarray}
\sigma_{L,T}^{\gamma^{*}p}(x,Q^{2})=\int{d^{2}\mathbf{r}}\int
dz\psi^{*}(Q,r,z) \hat{\sigma}(x,r)\psi(Q,r,z)
\end{eqnarray}
Here the quark (or antiquark) carries a fraction $z$ of the
incoming photon light-cone energy ($0<z<1$). The wave function
squared of the $q\overline{q}$ Fock states of the  virtual photon
is given by the following equations
\begin{eqnarray}
|\Psi_{T}(z,r)|^{2}&=&\frac{6\alpha_{em}}{4\pi^{2}}\sum_{1}^{n_{f}}e_{f}^{2}\{[z^{2}+(1-z)^{2}]
\epsilon^{2}K_{1}^{2}(\epsilon{r})\nonumber\\
&&+m^{2}_{f}K_{1}^{2}(\epsilon{r})\},\nonumber\\
\mathrm{and}~~~~~~~~~\nonumber\\
|\Psi_{L}(z,r)|^{2}&=&\frac{6\alpha_{em}}{4\pi^{2}}\sum_{1}^{n_{f}}e_{f}^{2}\{4Q^{2}z^{2}(1-z)^{2}K_{0}^{2}(\epsilon{r})\}
\end{eqnarray}
where $\epsilon^{2}=z(1-z)Q^{2}+m_{f}^{2}$ and $m_{f}$ is the
quark mass. $e_{f}$ is the quark charge and the functions
$K_{0,1}$ are the Bessel-McDonald functions. With respect to the
GBW model [11] inspired by geometrical scaling, the dipole
cross-section is defined by
\begin{eqnarray}
\sigma_{\mathrm{dipole}}(x,Q^{2})=\sigma_{0}\{1-\exp(-\frac{r^{2}Q^{2}_{s}(x)}{4})\}.
\end{eqnarray}
Authors in Ref.[24] defined the cross-section of $\gamma^{*}p$
scattering by the following form
\begin{eqnarray}
\sigma_{L,T}^{\gamma^{*}p}(x,Q^{2})&=&-\sigma_{0}\int\mathcal{F}_{L,T}
(\zeta,\frac{m_{f}^{2}}{Q^{2}})[\frac{Q^{2}_{0}}{Q^{2}}(\frac{x}{x_{0}})^{-\lambda}]^{1/2+i\zeta}\nonumber\\
&&{\times}\Gamma(-\frac{1}{2}-i\zeta)\frac{d\zeta}{2\pi}
\end{eqnarray}
Therefore the massive analytic proton structure function reads
\begin{eqnarray}
F_{2}(x,Q^{2})&=&\frac{Q^{2}}{4\pi^{2}\alpha}\sigma'_{0}\bigg{\{}\ln\bigg[(\frac{x_{0}}{x})^{\lambda}
\frac{Q_{0}^{2}}{Q^{2}+4m^{2}_{f}}+1\bigg]\nonumber\\
&&+(\frac{x_{0}}{x})^{\lambda}
\frac{Q_{0}^{2}}{Q^{2}+4m^{2}_{f}}\nonumber\\
&&{\times}\ln\bigg[(\frac{x}{x_{0}})^{\lambda}
\frac{Q^{2}+4m^{2}_{f}}{Q_{0}^{2}}+1\bigg] \bigg{\}},
\end{eqnarray}
where the fixed parameters in this relation determined in Ref.[24]
by fitting to small $x$ experimental data of the proton structure
function. Indeed Eq. (21) become to massless structure function
when $m_{f}$ sets to light quarks ($m_{f}=0.140~\mathrm{GeV}$),
and become to massive structure function when
$m_{f}{\rightarrow}m_{\mathcal{Q}}$.\\
In a convenient model [26], the cross-section of $\gamma^{*}p$
scattering  as function of the scaling variable
$\tau=Q^{2}/Q^{2}_{s}$ is proposed. A single universal curve,
suggested by N. Armesto, C. Salgado, and U.A. Wiedemann [27], in
what follows referred to as the ASW model (ASWM), describes the
available experimental data on $\sigma^{\gamma^{*}p}$ by the
following form
\begin{eqnarray}
\sigma^{\gamma^{*}p}(x,Q^{2}){\equiv}\Phi(\tau)=
\overline{\sigma}_{0}[\gamma_{E}+\Gamma(0,\xi)+{\ln}\xi]
\end{eqnarray}
where $\gamma_{E}$ and $\Gamma(0,\xi)$ are the Euler constant and
the incomplete $\Gamma$ function, respectively. Authors in
Ref.[27] extracted the $\xi$ function from a fit to lepton-proton
data as $\xi=a/\tau^{b}$ with $a=1.868$ and $b=0.746$ for massless
flavors. Therefore the proton structure function in ASW model
reads
\begin{eqnarray}
F_{2}^{ASW}(x,Q^{2})=\frac{Q^{2}}{4\pi^{2}\alpha}\overline{\sigma}_{0}
\bigg[ \gamma_{E}+\Gamma(0,\xi)+{\ln}\xi \bigg].
\end{eqnarray}
The ASW structure function (i.e., Eq.23) is considered  in
saturation physics.\\

\subsection{4. Heavy quarks structure functions}

The standard parameterization of the singlet and gluon
distribution functions with assuming the Regge-like behavior at
small $x$ is given by [28]
\begin{eqnarray}
F_{2}^{s}(x,Q^{2})_{x{\rightarrow}0}&=&x^{-\delta}\widetilde{s}(x,Q^{2}),\nonumber\\
xf_{g}(x,Q^{2})_{x{\rightarrow}0}&=&x^{-\delta}\widetilde{g}(x,Q^{2}).
\end{eqnarray}
The Regge-like behavior of the distribution functions was proposed
by Lopez and Yndurain [29]. The inclusive electroproduction on a
proton was studied at low $x$ and low $Q^{2}$ using a soft and
hard Pomeron in Ref.[30]. The value obtained by fixed coupling LLx
BFKL gives $\delta{\simeq}0.5$, which is the so-called
hard-Pomeron exponent. Some other phenomenological models have
also been proposed for the singlet structure function exponent in
Refs.[10] and [31]. In Ref.[32] authors presented a tensor-Pomeron
model where it is applied to low-$x$ deep inelastic lepton-nucleon
scattering and photoproduction processes. In this model, in
addition to the soft tensor Pomeron, a hard tensor Pomeron and
Reggeon exchange included. An effective behavior for the singlet
structure function is reported in Refs.[10] and [33]. This
effective exponent was found to be independent of $x$ and to
increase linearly with ${\ln}Q^{2}$. In Ref.[34] the CTEQ6.6C4
parametrization, which has a strong sea-like intrinsic charm
contribution, give the $\delta$ value that is the closet to $0.3$
found in data and leads to a much better quality factor (QF) than
the other parametrizations (as reported in Ref.[15]). For the
behavior of the distribution functions and saturation scale, we
use the GBW parametrization [11] with $\delta=0.288$. This
exponent of the saturation scale is known at next-to-leading order
[35] and agrees with the values extracted from fits to new combined data.\\
Now exploiting the low $x$ asymptotic behavior of
$f_{s,g}(x,Q^{2})$ according to Eq.(24). We obtain the following
equations for the $Q^{2}$ derivative of the structure function
(i.e., Eq.(14)) and heavy quark structure functions (i.e.,
Eq.(11)) by the following forms
\begin{eqnarray}
\frac{{\partial}F_{2}(x,Q^{2})}{{\partial}{\ln}Q^{2}}&=&-\frac{1}{2}[
r_{ss}(x,a_{s},\delta)F_{2}(x,Q^{2})\nonumber\\
&&+<e^{2}>r_{sg}(x,a_{s},\delta)xf_{g}(x,Q^{2})],\nonumber\\
F_{k}^{\mathcal{Q}\overline{\mathcal{Q}}}(x,Q^{2})&=&r_{k,g}(x,a_{s},\delta)xf_{g}(x,Q^{2}),
\end{eqnarray}
where
\begin{eqnarray}
r_{ab}(x,a_{s},\delta)&=& \sum_{n=0}\widetilde{P}^{(n)}_{ab}(x,a_{s}){\odot}x^{\delta},\nonumber\\
r_{k,g}(x,a_{s},\delta)&=&
\sum_{n=0}C_{k,g}^{(n)}(x,a_{s}){\odot}x^{\delta},\nonumber
\end{eqnarray}
where $r^{,}s$ is defined due to the splitting and Wilson
coefficient functions. The convolution form $A(x){\odot}B(x)$
reads to be [36]
$$
A(x){\odot}B(x){\equiv}\int\frac{dz}{z}A(a_{s},z)B(z).
$$
From Eqs.(25) one can obtain the heavy quark structure functions
as a function of $F_{2}$ and the derivative as
\begin{eqnarray}
F_{k}^{\mathcal{Q}\overline{\mathcal{Q}}}(x,Q^{2})&=&-\frac{1}{<e^{2}>}\frac{r_{k,g}(x,a_{s},\delta)}{r_{sg}(x,a_{s},\delta)}
[2\frac{{\partial}F_{2}(x,Q^{2})}{{\partial}{\ln}Q^{2}}\nonumber\\
&&+r_{ss}(x,a_{s},\delta)F_{2}(x,Q^{2})].
\end{eqnarray}
Using the coefficient functions in the NNLO approximation for
concrete value of $\delta=0.288$ we obtain an analytical equation
for the heavy quark structure functions from above formulae (26)
in the arguments of the function $F_{2}$ and its derivative due to
the
BDH and ASW models.\\
A particular interests present the ratio of the longitudinal to
transversal heavy quark structure functions, defined as
\begin{eqnarray}
R^{\mathcal{Q}\overline{\mathcal{Q}}}(x,Q^{2})=\frac{F_{L}^{\mathcal{Q}\overline{\mathcal{Q}}}(x,Q^{2})}{F_{2}^{\mathcal{Q}\overline{\mathcal{Q}}}(x,Q^{2})}
=\frac{r_{L,g}(x,a_{s},\delta) }{r_{2,g}(x,a_{s},\delta) }.
\end{eqnarray}
In fact the proton structure function and its derivative cancel in
the ratio of heavy quark structure functions, which is very useful
for practical applications.\\

\subsection{4. RESULTS}

With help of Eqs.(26) and (27) we have extracted the heavy  quark
structure functions and the ratio
$R^{\mathcal{Q}\overline{\mathcal{Q}}}$
 from the  parametrizations  of the proton structure
functions (i.e., BDH and ASW models) respectively. As for our
input parameters, we choose the running charm and bottom quark
masses $m_{c}=1.29^{+0.077}_{-0.053}~\mathrm{GeV}$ and
$m_{b}=4.049^{+0.138}_{-0.118}~\mathrm{GeV}$, where the
uncertainties are obtained through adding the experimental fit,
model and parameterization uncertainties in quadrature [7]. The
strong coupling constant value  is chosen to be
$\alpha_{s}(M_{z}^{2})=0.118$. We employ the standard
representation for QCD coupling and the coefficient functions  in
the NNLO approximation. To investigate GS for heavy quark
structure functions we shall compare results obtained using
scaling variable $\tau$ (8). The parameters $Q_{0}$ and $x_{0}$
are related to the GBW model by $Q_{0}=1~\mathrm{GeV}$ and
$x_{0}=3.04{\times}10^{-4}$. Let us now confront the implications
of geometrical scaling with our results and experimental data on
heavy quark structure functions at low $x$. We have calculated the
$\tau$-dependence, at low $x$, of the heavy quark structure
functions $F_{2}^{\mathcal{Q}\overline{\mathcal{Q}}}(\tau)/\tau$
with respect to the parametrization of $F_{2}^{\mathrm{BDH}}$ and
$F_{2}^{\mathrm{ASW}}$ as described above, in the NNLO
approximation, Eq. (26). Results of calculations and comparison
with HERA data [16] for charm and bottom structure functions are
presented in Figs.1 and 2, where the circles correspond to the
extracted charm and bottom structure functions due to the BDH and
ASW models in the NNLO approximation, respectively.\\
\begin{figure}[h]
\includegraphics[width=0.5\textwidth]{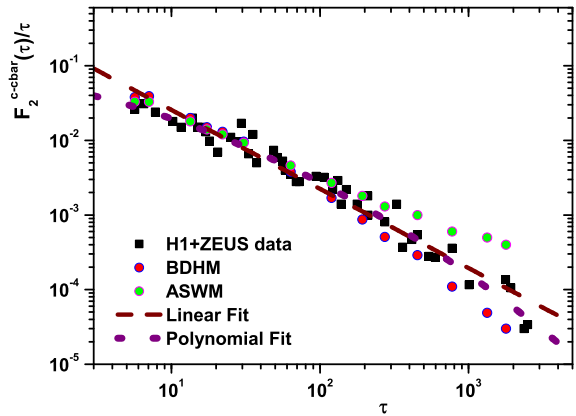}
\caption{ $F_{2}^{c\overline{c}}(\tau)/\tau$ as a function of
scaling variable $\tau$ with $m_{c}=1.29$. Two circles correspond
to proton structure functions due to the BDH and ASW models.
Experimental data on the charm structure functions from the region
$x<0.1$ collected corresponding to the H1 2010 and ZEUS 2014 data
[16]. The linear (dashed curve) and non-linear (dotted curve) fits
plotted in comparison with the HERA data. }\label{Fig1}
\end{figure}
\begin{figure}[h]
\includegraphics[width=0.5\textwidth]{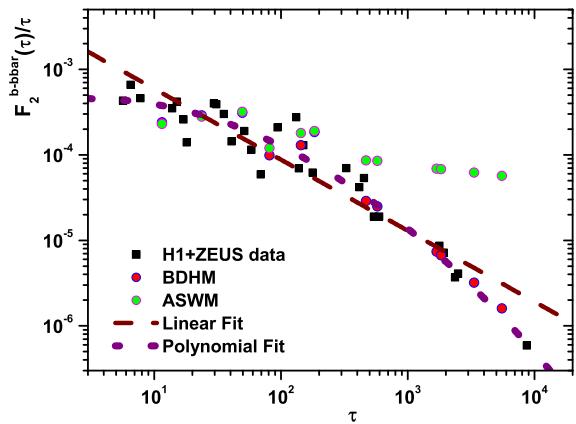}
\caption{ $F_{2}^{b\overline{b}}(\tau)/\tau$ as a function of
scaling variable $\tau$ with $m_{b}=4.049$. Two circles correspond
to proton structure functions due to the BDH and ASW models.
Experimental data on the bottom structure functions from the
region $x<0.1$ collected corresponding to the H1 2010 and ZEUS
2014 data [16]. The linear (dashed curve) and non-linear (dotted
curve) fits plotted in comparison with the HERA data.}\label{Fig2}
\end{figure}
Figures 1 and 2 clearly demonstrate that the extraction procedures
provide correct behaviors of the extracted heavy quark structure
functions in both models at small $\tau$. At large $\tau$ the
extracted heavy quark structure functions due to the BDHM are in a
good agreement with experimental data. This is visible as the ASWM
has good agreement with the proton structure function for
$\tau<100$ [26]. The deviation between these curves therefore
shows the effect of the heavy quarks masses. In Figs.1 and 2 we
show results for $F_{2}^{c\overline{c}}(\tau)/\tau$ and
$F_{2}^{b\overline{b}}(\tau)/\tau$ as a function of the scaling
variable $\tau$ for different $Q^{2}$ above $5~\mathrm{GeV}^{2}$
and $x{\geq}0.0002$ due to the data  of the H1-Collaboration [16]
and for $6.5~\mathrm{GeV}^{2}$ and $x{\geq}0.00015$ due to the
data of the ZEUS-Collaboration [16]. The theoretical results are
presented for the same kinematical variables as the experimental
data for average of the Bjorken values of $x$, and we see that
there are  good agreements between  BDH theory and data at small
and  large $\tau$.\\
In Fig.1 we see that data exhibit GS over a very broad region of
$\tau$. We can clearly see that the charm structure functions
reflects the fact that $F_{2}^{c\overline{c}}(\tau)/\tau$ scales
as $1/\tau$. The linear fit shows this scaling with a good
accuracy. However a second-order polynomial fit shows that the
change of shape of the dependence of
$F_{2}^{c\overline{c}}(\tau)/\tau$ on $\tau$ from the approximate
$1/\tau$ dependence at large $\tau$ to the less steep dependence
at small $\tau$ [8,37]. Important result is that the linear and
non-linear fits to the charm structure functions exhibits GS by
the following forms
\begin{eqnarray}
F_{2}^{c\overline{c}}(\tau)|_{\mathrm{Non-Linear}}&{\simeq}&0.06~\tau^{0.69}\exp[-0.08~{\ln^{2}}(\tau)],\nonumber\\
F_{2}^{c\overline{c}}(\tau)|_{\mathrm{Linear}}&{\simeq}&0.30~\tau^{-0.06}.
\end{eqnarray}
The difference between linear and non-linear curves therefore
represents the effect of saturation. From Fig.2 one can infer that
the non-linear results essentially improve the agreement with data
in comparison with the linear fit. We see that the experimental
data are very well reproduced by the non-linear fits for bottom
structure functions, which gives the behavior of
$F_{2}^{b\overline{b}}(\tau)/\tau$ on $\tau$ from the asymptotic
$1/\tau$ dependence at large $\tau$ to the less steep dependence
at small $\tau$, which corresponds to the fact that at small
values of $\tau$ the bottom structure function grows weaker with
energy than $Q^{2}_{s}(x)$. The linear and non-linear fits to the
bottom structure functions are given by
\begin{eqnarray}
F_{2}^{b\overline{b}}(\tau)|_{\mathrm{Non-Linear}}&{\simeq}&0.0004~\tau^{1.2300}\exp[-0.1040~{\ln^{2}}(\tau)],\nonumber\\
F_{2}^{b\overline{b}}(\tau)|_{\mathrm{Linear}}&{\simeq}&0.004~\tau^{0.172}.
\end{eqnarray}
In Figs.(3) and (4) we also found a symmetry between the regions
of large and small $\tau$ for the heavy quark structure functions
due to the non-linear fits in comparison with the HERA data [16].
\begin{figure}[h]
\includegraphics[width=0.5\textwidth]{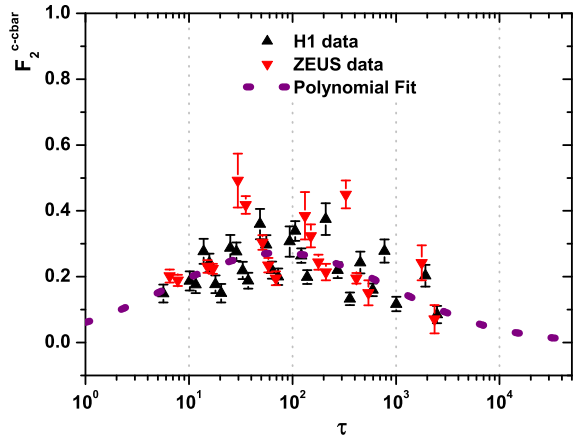}
\caption{ The non-linear fits to the charm structure functions
$F_{2}^{c\overline{c}}(\tau)$ plotted versus the scaling variable
$\tau$. Experimental data are from the H1  and ZEUS
Collaborations, Ref.[16] as accompanied with total errors.
}\label{Fig3}
\end{figure}
\begin{figure}[h]
\includegraphics[width=0.5\textwidth]{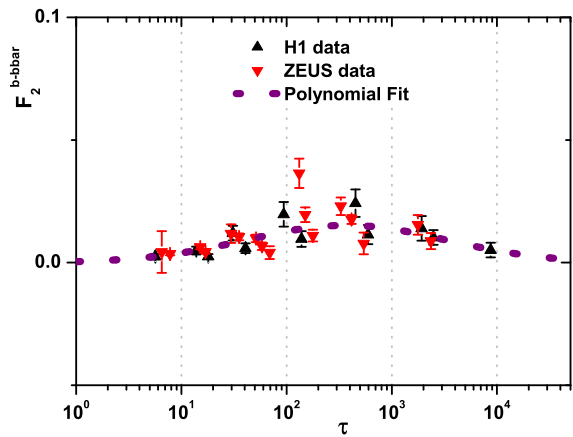}
\caption{ The non-linear fits to the bottom structure functions
$F_{2}^{b\overline{b}}(\tau)$ plotted versus the scaling variable
$\tau$. Experimental data are from the H1  and ZEUS
Collaborations, Ref.[16] as accompanied with total errors.
}\label{Fig4}
\end{figure}
Figs.(3) and (4) indicate the present of symmetry of
$F_{2}^{c\overline{c}}$ and $F_{2}^{b\overline{b}}$ in comparison
with the HERA data [16] with respect to the transformation
$\tau{\leftrightarrow}1/\tau$ in a wide range of $\tau$ values
[38]. With respect to Figs.(3) and (4), we also found a symmetry
line between the regions of large and small $\tau$ for the charm
and bottom structure functions, which corresponds to the ratio of
heavy quarks masses, $m_{b}/m_{c}{\simeq}3.139$. The argument in
symmetry line is $\tau_{b}{\simeq}3.139~\tau_{c}$. In conclusion
the features present in the average of the charm and bottom
structure function can be well reproduced in the phenomenological
saturation model as shown in (28) and (29), corresponding to the
non-linear dashed curves in Figs.3 and 4.
\begin{figure}[h]
\includegraphics[width=0.5\textwidth]{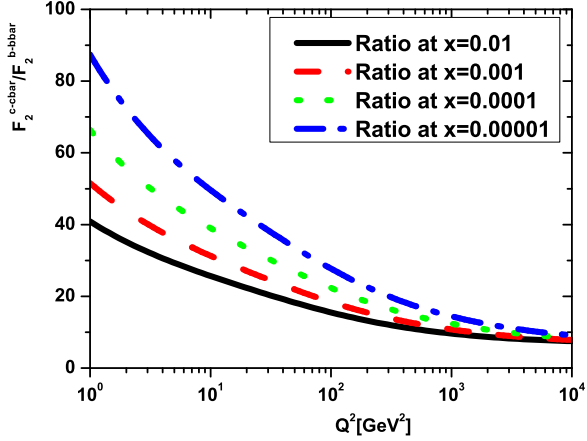}
\caption{ $F_{2}^{c\overline{c}}/F_{2}^{b\overline{b}}$ for fixed
$x$ as a function of $Q^{2}$. }\label{Fig5}
\end{figure}
\begin{figure}[h]
\includegraphics[width=0.5\textwidth]{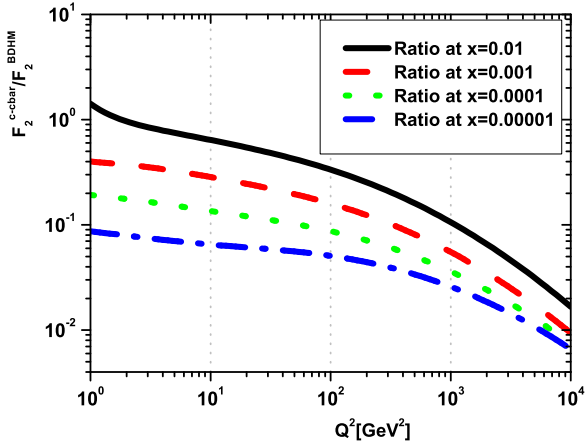}
\caption{ The ratio $F_{2}^{c\overline{c}}/F_{2}^{BDHM}$ is
extracted from the charm non-linear structure function (i.e.,
Eq.28) and the parametrization of the proton structure function
(i.e., Eq.15), in a wide range of $Q^{2}$ for $x=10^{-2..-5}$.
}\label{Fig6}
\end{figure}
In Fig.5 we show comparison of the heavy quark structure
functions, $F_{2}^{c\overline{c}}/F_{2}^{b\overline{b}}$, obtained
using the non-linear behaviors for $x=10^{-2..-5}$. The ratio of
the heavy quarks structure function  increase as $x$ decreases at
low $Q^{2}$. It is clear that at low $Q^{2}$ values the FFNS
approach is valid. Here heavy flavors are not considered as active
and are generated only by boson-gluon fusion (BGF). The charm and
bottom quarks are infinity massive for
$Q^{2}{\leq}m^{2}_{\mathcal{Q}}$ and are massless above this
threshold. For $Q^{2}{\gg}m^{2}_{\mathcal{Q}}$  NNLO terms in the
coefficient functions are enhanced by powers of
$\log(Q^{2})/m^{2}_{\mathcal{Q}}$ where VFNS is valid in this
region [15]. As we observe in Fig.5, this ratio is expected to
rise for a given $Q^{2}$ with decreasing $x$ and fall for a given
$x$ with increasing $Q^{2}$, respectively. Indeed the ratio
$F_{2}^{c\overline{c}}/F_{2}^{b\overline{b}}$ at large $Q^{2}$ is
almost constant as for $Q^{2}=10^{4}~\mathrm{GeV}^{2}$ the ratio
is independent of the $x$ variable (illustrated in right-hand side
of Fig.5).
\begin{figure}[h]
\includegraphics[width=0.5\textwidth]{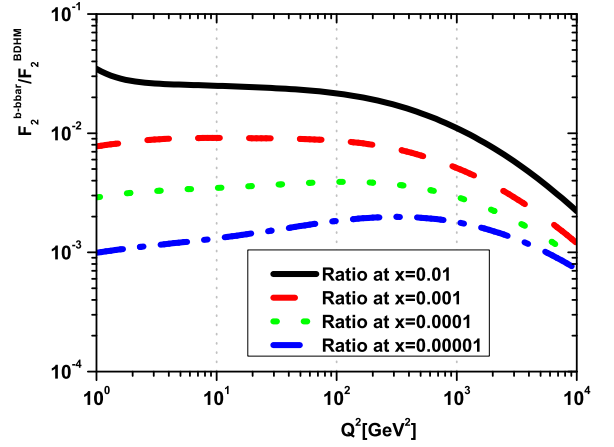}
\caption{ The ratio $F_{2}^{b\overline{b}}/F_{2}^{BDHM}$ is
extracted from the bottom non-linear structure function (i.e.,
Eq.29) and the parametrization of the proton structure function
(i.e., Eq.15), in a wide range of $Q^{2}$ for $x=10^{-2..-5}$.
}\label{Fig7}
\end{figure}
In Figs.6 and 7 we  investigated the  ratio
$F_{2}^{c\overline{c}}/F_{2}^{BDHM}$ and
$F_{2}^{b\overline{b}}/F_{2}^{BDHM}$ whose behavior is very
interesting in the range of available HERA energy [39,40] and its
extension to future energies in LHeC and FCC-eh [1]. In these
figures (i.e., Figs.6 and 7) we shown the ratio of the heavy
quarks structure functions where the $F_{2}(x,Q^{2})$
parameterization is taken from Ref.[22]. It is interesting that
the values of  these ratios are constant due to the HERA data
range. HERA data range on the cross sections for the open charm
and bottom production in neutral current deep inelastic
electron-proton scattering (DIS) extended from
$Q^{2}{\sim}\mathcal{O}(5~\mathrm{ GeV}^{2})$ and $x\sim10^{-4}$
until $Q^{2}{\sim}\mathcal{O}(1000~\mathrm{ GeV}^{2})$ and
$x\sim10^{-2}$. Therefore in Fig.6 we observe that the ratio
$F_{2}^{c\overline{c}}/F_{2}^{BDHM}$ is approximately between
$\sim0.2-0.1$ as this prediction is close to the average result
$<F_{2}^{c\overline{c}}/F_{2}>= 0.237{\pm}0.021{\pm}0.041$, in
Ref.[39] and it is compatible with the measured results by EMC. In
Fig.7 the ratio $F_{2}^{b\overline{b}}/F_{2}^{BDHM}$ with respect
to the HERA region is approximately between $\sim0.003-0.01$. It
is interesting that with increasing $x$ and $Q^{2}$ according to
the HERA region, the ratio $F_{2}^{c\overline{c}}/F_{2}^{BDHM}$
decreases while the ratio $F_{2}^{b\overline{b}}/F_{2}^{BDHM}$
increases. These results in a wide range of $x$ and $Q^{2}$ can be
applied as well in analyses of ultrahigh energy processes with
cosmic neutrinos and will be able to be considered in the LHeC and
FCC-eh collisions. In Figs.8 and 9  the photon-proton cross
sections for charm and bottom production are shown, as
$\sigma^{c\overline{c}}{\sim}F_{2}^{c\overline{c}}/Q^{2}$ and
$\sigma^{b\overline{b}}{\sim}F_{2}^{b\overline{b}}/Q^{2}$. In
these figures we tested the properties of
$F_{2}^{\mathcal{Q}\overline{\mathcal{Q}}}/Q^{2}$ using the
parameterization of the heavy-quark structure functions in a wide
range of $x$ and $Q^{2}$. The scaling properties of these
functions as a function of the scaling variable $\tau$ are shown
in Figs.1 and 2. Geometrical scaling in heavy-quarks production
should predominantly be regarded as a remarkable regularity of the
inclusive DIS.
\begin{figure}[h]
\includegraphics[width=0.5\textwidth]{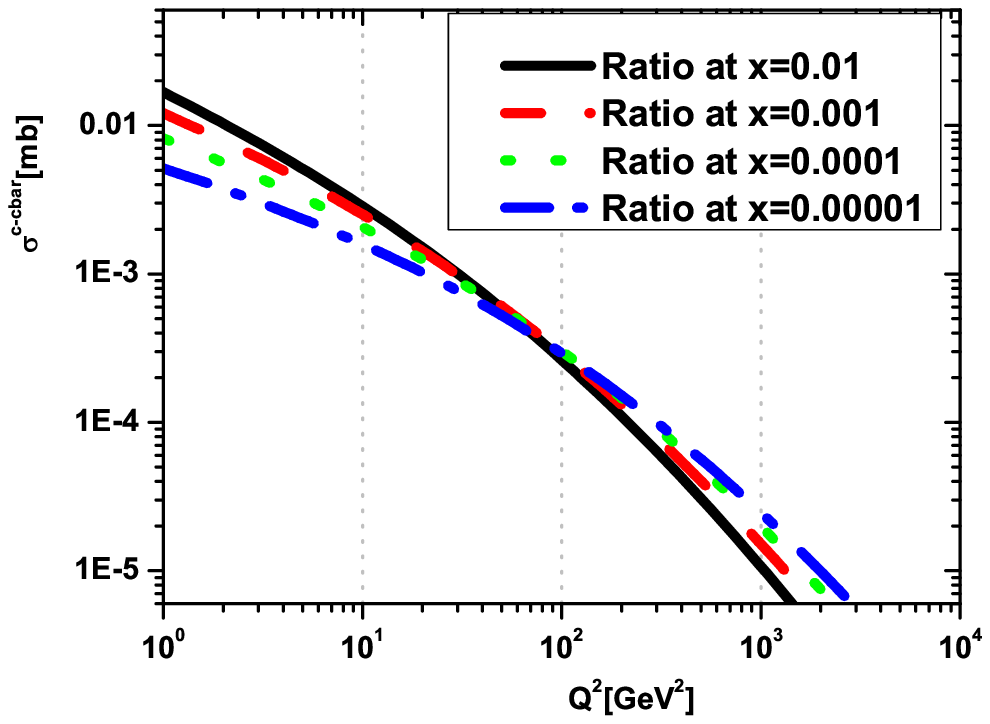}
\caption{ The charm component
$\sigma^{c\overline{c}}{\sim}F_{2}^{c\overline{c}}/Q^{2}$ in a
wide range of $Q^{2}$ for $x=10^{-2..-5}$. }\label{Fig8}
\end{figure}
\begin{figure}[h]
\includegraphics[width=0.5\textwidth]{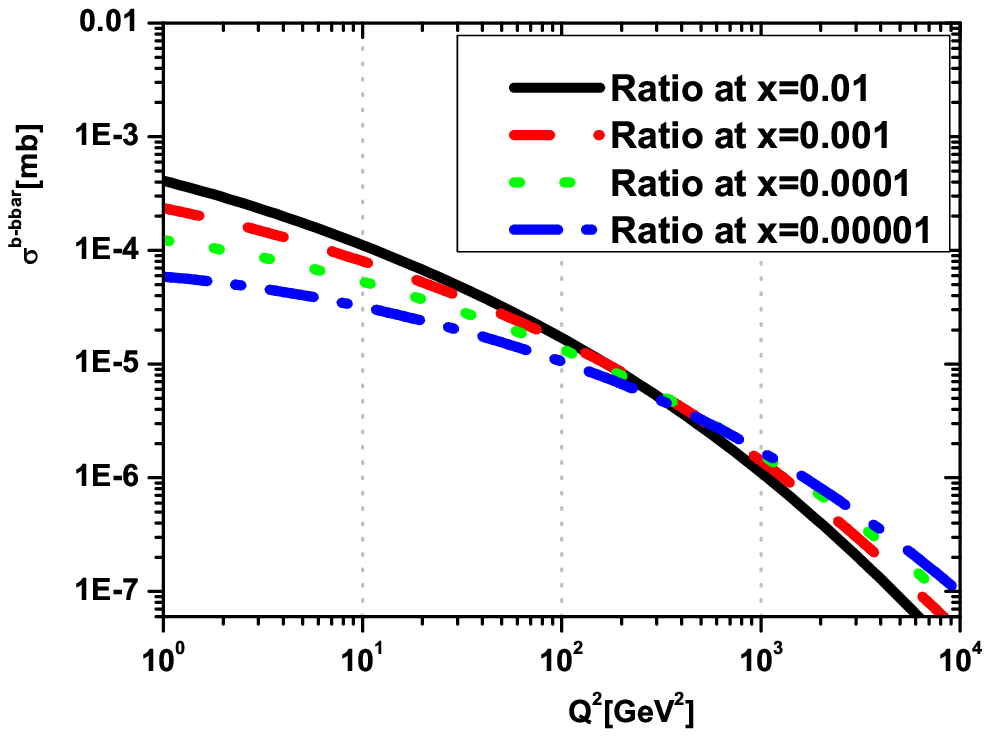}
\caption{ The bottom component
$\sigma^{b\overline{b}}{\sim}F_{2}^{b\overline{b}}/Q^{2}$ in a
wide range of $Q^{2}$ for $x=10^{-2..-5}$. }\label{Fig9}
\end{figure}
In Fig.10 we show  the $\tau$ dependence of $R^{c}(\tau)$
evaluated at NNLO approximation from Eq.(27) at
$<\mu^{2}>=Q^{2}+2m^{2}_{c}$. Our calculations in the conventional
(collinear) QCD factorization leads to more or less flat
(independent on $\tau$) behavior of $R^{c}(\tau)$ with
$0.1~{\lesssim}~R^{c}(\tau)<0.2$ in a wide range of $\tau$. The
results obtained are compatible with the collinear results in DAS
approach and the TMD results at low $Q^{2}$. In Refs.[40-45] the
results for the ratio of heavy quarks structure functions,
obtained using the power-like behavior $x^{-\delta}$, can be
found. Our calculations show that the collinear DAS  predictions
rather close to the results obtained due to the geometrical
scaling at NNLO approximation. Indeed, the NNLO results obtained
in the collinear perturbation theory due to the geometrical
scaling lead to small values for the ratio $R^{c}(\tau)$ in
comparison with the corresponding $Q^{2}$-dependence evaluated
with the TMD gluons. In fact, in collinear perturbation theory the
ratio $R^{c}(\tau)$ grow slowly when $\tau$ increased.
\begin{figure}[h]
\includegraphics[width=0.5\textwidth]{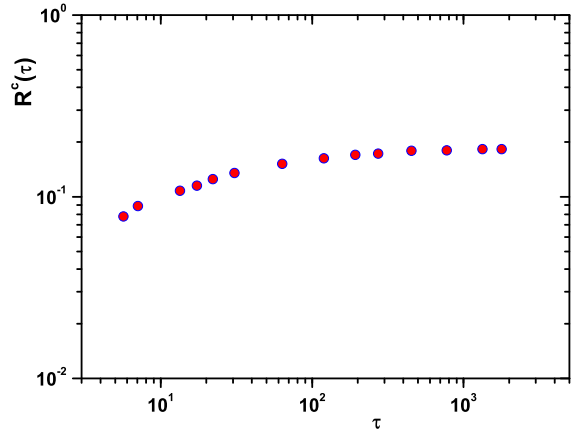}
\caption{ The ratio of the charm structure function $R^{c}(\tau)$
as a function of $\tau$. }\label{Fig10}
\end{figure}

\subsection{5. CONCLUSIONS}

We have shown that the charm and bottom structure function data
exhibit geometrical scaling at low $x$ in a wide range of $Q^{2}$,
where $F_{2}^{\mathcal{Q}\overline{\mathcal{Q}}}(\tau)/\tau$ is
the function of only one dimensionless variable
$\tau=Q^{2}/Q^{2}_{sat}(x)$ with taking into account charm and
bottom masses. We have used data from the HERA experiment [16].
Results obtained there suggest that polynomial fits predict the
geometrical scaling of the charm and bottom structure function
with a good accuracy. At low $\tau$, geometrical scaling may
provide genuine evidence for parton saturation. We see the heavy
quark structure functions exhibit geometric scaling for the whole
$Q^{2}$ range, verifying a transition in the behavior on $\tau$ of
the heavy quark structure functions from a smooth dependence at
small $\tau$ and an approximated $1/\tau$ behavior at large
$\tau$. The transition point between the charm and bottom
structure functions is placed at $<\mu^{2}>$ , which takes values
of order $m_{b}/m_{c}$ for a charm and bottom masses
$m_{c}=1.290~\mathrm{GeV}$ and $m_{b}=4.049~\mathrm{GeV}$
respectively, and  found a symmetry between the regions of large
and small $\tau$ for the charm and bottom structure functions with
respect to the transformation $\tau{\leftrightarrow}1/\tau$ in the
whole region of $\tau$.\\
In fact we have presented Eq.(26) for the extraction of the heavy
quark structure function
$F_{2}^{\mathcal{Q}\overline{\mathcal{Q}}}$ at low $x$ from the
$F_{2}$ and its $\ln{Q^{2}}$ derivative with the explicit form of
the proton structure function by the BDH and ASW parametrizations
using the DAS approach at NNLO approximation. Moreover, we
obtained the results for
$F_{2}^{c\overline{c}}/F_{2}^{b\overline{b}}$,
$F_{2}^{c\overline{c}}/F_{2}^{BDHM}$ and
$F_{2}^{b\overline{b}}/F_{2}^{BDHM}$ where they are seems to be
extremely important for future colliders. Concerning the ratio
$R^{c}(\tau){\equiv}F_{L}^{c\overline{c}}/F_{2}^{c\overline{c}}$,
we show that the results are flat in a wide range of $\tau$ and
they are similar to the ones obtained beyond LO of
$k_{T}$-factorization calculations at low $\tau$ and of collinear
perturbation theory at large $\tau$. In Ref.[18], the
$k_{T}$-factorization predictions obtained using derived
analytical expressions for TMD gluon density. Finally we have
investigated the scaling properties of
$\sigma^{\mathcal{Q}\overline{\mathcal{Q}}}{\sim}F_{2}^{\mathcal{Q}\overline{\mathcal{Q}}}/Q^{2}$
with respect the parametrization of heavy quark structure function
at low $x$.\\

\subsection{ACKNOWLEDGMENTS}
The author is thankful to the Razi University for financial
support of this project.

\section{References}

1. LHeC Collaboration and FCC-he Study Group, P. Agostini et al., J. Phys. G: Nucl. Part. Phys. {\bf48}, 110501(2021).\\
2. S. Alekhin, J.Bl$\ddot{\mathrm{u}}$mlein and  S. Moch,
Phys.Rev.D {\bf102}, 054014 (2020).\\
3. J.Bl$\ddot{\mathrm{u}}$mlein et al., Nucl.Phys.B {\bf755}, 272
(2006).\\
4. S. Alekhin, J.Bl$\ddot{\mathrm{u}}$mlein and S. Moch, Phys.
Rev. D {\bf86}, 054009
(2012).\\
5. R.D. Ball et al. [NNPDF Collaboration], Nucl. Phys. B {\bf855},
153 (2012).\\
6. R.Thorne, Phys.Rev.D {\bf86}, 074017 (2012).\\
7. H.Abramowicz et al., [H1 and ZEUS Collaboration], Eur.Phys.J.C
{\bf78}, 473 (2018).\\
8. A.M.Stasto, K.Golec-Biernat and J.Kwiecinski, Phys.Rev.Lett.
{\bf86}, 596 (2001).\\
9. F.Gaola and S.Forte, Phys.Rev.Lett.
{\bf101}, 022001 (2008).\\
10. M.Paraszalowicz and T.Stebel, JHEP {\bf04}, 169 (2013);
T.Stebel, Phys.Rev.D {\bf88}, 014026 (2013).\\
11. K.Golec-Biernat and M.W$\ddot{\mathrm{u}}$sthoff, Phys.Rev.D
{\bf59}, 014017 (1998); Phys.Rev.D {\bf60}, 114023 (1999).\\
12. J.Jalilian-Marian, A.Kovner, A.Leonidov and H.Weigert,
Nucl.Phys.B {\bf504}, 415 (1997); Phys.Rev.D {\bf59}, 014014
(1999); E.Iancu, A.Leonidov and L.D.McLerran, Nucl.Phys.A
{\bf692}, 583 (2001); E.Ferreiro, E.Iancu, A.Leonidov and
L.D.McLerran, Nucl.Phys.A {\bf703}, 489 (2002).\\
13. L.D. McLerran and R. Venugopalan, Phys. Rev. D {\bf49}, 2233
(1994); Phys. Rev. D {\bf49}, 3352 (1994); A. Ayala, J.
Jalilian-Marian, L.D. McLerran and R. Venugopalan, Phys.Rev.D
{\bf52}, 2935 (1995); Phys.Rev.D {\bf 53}, 458 (1996).\\
14. A. Morreale and F. Salazar, Universe {\bf7}, 312 (2021).\\
15. G.Beuf, C.Royon and D.Salek, arXiv [hep-ph]:0810.5082; R.S.Thorne, 9805298.\\
16. F.D.Aaron et al., (H1 Collaboration), Eur.Phys.J.C {\bf 65},
89 (2010); H.Abramowicz et al., [ZEUS Collaboration], JHEP09,
127(2014).\\
17.  E. Laenen et al., Phys. Lett. B{\bf291}, 325 (1992); S.
Alekhin and S. Moch,
Phys. Lett. B{\bf699}, 345 (2011).\\
18. A.V.Kotikov, A.V.Lipatov and P.Zhang, Phys.Rev.D {\bf104},
054042 (2021).\\
19. M.Gluck, E.Reya and A.Vogt, Z.Phys.C {\bf67}, 433 (1998);
Eur.Phys.J.C {\bf5}, 461 (1995).\\
20. A.Yu.Illarionov, B.A.Kniehl and A.V.Kotikov, Phys.Lett.B
{\bf663}, 66 (2008).\\
21. H.Kawamura, N.A.Lo Presti, S.Moch and A.Vogt, Nucl.Phys.B
{\bf864}, 399 (2012).\\
22. M. M. Block, L. Durand and P. Ha, Phys. Rev.D {\bf 89}, 094027 (2014).\\
23. L.P.Kaptari, A.V.Kotikov, N.Yu.Chernikova and Pengming Zhang,
Phys.Rev.D {\bf99}, 096019 (2019); G.R.Boroun and B.Rezaei,
Phys.Rev.D {\bf105}, 034002 (2022).\\
24. Y.Hu et al., Eur.Phys.J.A {\bf51}, 159 (2015).\\
25. N.N.Nikolaev, B.G.Zakharov, Z.Phys.C {\bf49}, 607 (1990).\\
26. N.Armesto, C.Merino, G.Parente, E.Zas, Phys.Rev.D {\bf77},
013001 (2008).\\
27. N.Armesto, C.Salgado, and U.A.Wiedemann, Phys.Rev.Lett.
{\bf94}, 022002 (2005); J.L.Albacete, N.Armesto, J.G.Milhano,
C.A.Salgado, and U.A.Wiedemann, Eur.Phys.J.C {\bf43}, 353 (2005).\\
28. A.V.Kotikov, J.Exp.Theor.Phys. {\bf80}, 979 (1995);
A.V.Kotikov and G.Parente, Mod.Phys.Lett.A {\bf12}, 963 (1997).\\
29. C.Lopez, F.J.Yndurain, Nucl.Phys.B {\bf171},231 (1980).\\
30. J.R.Cudell, A.Donnachie and P.V.Landshoff, Phys.Lett.B
{\bf448}, 281 (1999).\\
31. B.Rezaei, G.R.Boroun, Eur.Phys.J.A {\bf55}, 66 (2019).\\
32. D.Britzger, et al., Phys.Rev.D {\bf100}, 114007 (2019).\\
33. C.Adloff, et al., H1 Collaboration, Phys.Lett.B {\bf520}, 183
(2001).\\
34. P.M.Nadolsky et al., Phys.Rev.D {\bf78}, 013004 (2008).\\
35. D.N.Triantafyllopoulos, Nucl.Phys.B {\bf648}, 293 (2003).\\
36. G.R.Boroun and B.Rezaei, Phys.Lett.B {\bf816}, 136274
(2021).\\
37. E.Avsar and G.Gustafson, JHEP{\bf04}, 067(2007).\\
38. V.P.Goncalves and M.V.T.Machado, Phys.Rev.Lett. {\bf91}, 202002(2003).\\
39. K.Daum et al., arXiv [hep-ph]:9609478.\\
40. G.R.Boroun and B.Rezaei, EPL {\bf133}, 61002 (2021).\\
41. N.N.Nikolaev, J.Speth and V.R.Zoller, Phys.Lett.B {\bf473},
157 (2000); N.N.Nikolaev and V.R.Zoller, Phys.Lett.B {\bf509},
283(2001); N.N.Nikolaev and V.R.Zoller, Phys.Atom.Nucl. {\bf73},
672(2010); R.Fiore, N.N.Nikolaev and V.R.Zoller, JETP Lett. {\bf90}, 319 (2009).\\
42. G.R.Boroun and B.Rezaei, Int.J.Mod.Phys.E {\bf24}, 1550063
(2015); Nucl.Phys.A {\bf929}, 119 (2014); Nucl.Phys.B {\bf 857},
143 (2012); EPL {\bf100}, 41001 (2012). \\
43. N.Ya.Ivanov, Nucl.Phys.B {\bf 814}, 142 (2009).\\
44. G.R.Boroun, Chin.Phys.C {\bf45}, 063105 (2021); Nucl.Phys.B {\bf 884}, 684 (2014)\\
45. A.Donnachie and P.V.Landshoff, Phys.Lett.B {\bf470}, 243
(1999).\\
\end{document}